\documentstyle[12pt,epsf]{article}
\vspace{-1cm}
\newcommand{\umu}{\mbox{$U_{\mu}(x)\ $} }
\newcommand{\amu}{\mbox{$A_{\mu}(x)\ $} }
\newcommand{\G}{\mbox{$G(x)\ $}}

\begin{document}
\renewcommand{\thefootnote}{\fnsymbol{footnote}}
\begin{titlepage}
\begin{center}
\Large\bf
A Comparative Study of Gauge Fixing Procedures\\
on the Connection Machines CM2 and CM5
\end{center}
\vspace{2cm}
\begin{center}
{\bf H. Suman\footnote[2]{E-mail:
suman@wptu23.physik.uni-wuppertal.de.} and K. Schilling\\
\em Fachbereich Physik, Bergische Universit\"at--Gesamthochschule
Wuppertal\\
Gau\ss str. 20, 5600 Wuppertal, Germany}
\end{center}
\vspace{5cm}
\centerline{\Large Abstract}
\begin{center}
\begin{minipage}[t]{12cm}
Gauge fixing is a frequent task encountered in practical lattice gauge 
theory calculations.
We review the performance characteristics of some standard gauging procedures 
for non-Abelian gauge theories, implemented on the parallel machines CM2
and CM5.
\end{minipage}
\end{center} 
\vfill
\end{titlepage}
\newpage 
\renewcommand{\thefootnote}{\arabic{footnote}}
\setcounter{footnote}{0}          
\section{Introduction}
Large-scale computer simulations have become an important tool in
the theory of elementary particles. There is a very rich empirical
material to test quantum chromodynamics (QCD), which is held
to be the basic theory of strong interactions\cite{qcd}.  Most of it 
is related to low energy data such as the spectrum and the structure of hadrons,
where nonperturbative methods such as lattice gauge theory (LGT) in
four dimnsions are indispensable\cite{lgt}.

In many practical situations of LGT it is necessary to perform gauge
fixing on the discrete lattice. This is in particular the
case if one is interested in the computation of gauge-noninvariant
quantities such as gluon propagators\cite{gp}\cite{mandgp}, Bethe-Salpeter
amplitudes\cite{bsa} or monopole densities\cite{mono}. Gauge fixing 
might also be helpful to reduce noise in the measurement of gauge 
invariant quantities (such as Polyakov lines or Wilson loops)\cite{inv}, 
or in the context of preconditioning\cite{dav}.

Various methods have been proposed to achieve gauge fixing
\cite{dav}\cite{gusor}. 
They are based on iterative schemes, which might need many
iteration steps and thus tend to be rather time consuming.
It is therefore of considerable interest to study their
convergence behaviour and the efficiency  of their implementations 
on parallel computers.

In this paper we will present an introduction into the issues and compare 
two  such algorithms for
Landau gauge fixing in lattice QCD,  on our local connection machines,
CM2 and CM5. We will also consider some  variants
of these basic algorithms, that aim at  possible
acceleration gains.

\section{Gauge fixing on the lattice}

QCD gauge fields are defined on the lattice in terms of parallel
transport operators, \umu $\in SU(3)$, $\mu = 1..4$,
that live on the  links between neighbouring sites $x$, $x+\hat{\mu} $ 
of a fourdimensional hypercubic lattice and are related 
to the continuum fields $A_{\mu}(x)$ by
\begin{equation}
A_{\mu}=\frac{1}{2i}\left( U_{\mu}(x)-U_{\mu}^+(x)\right) \mid_{traceless}.
\end{equation}
\umu are matrices in a threedimensional `color' space, and the
set \{$U_{\mu}$\} is called a configuration.
Under a local  gauge transformation \umu transforms like 
\begin{equation}
\umu\ \longrightarrow\ \umu^{(G)}=G(x)\umu G(x+\hat{\mu}),
\end{equation}
where the matrices G are elements of the  gauge group $ SU(3) $ and 
are associated to the lattice sites.

In the continuum formulation the {\it local} Landau gauge fixing
condition, $\partial_{\mu} A^{\mu}=0,$ can be viewed as the solution
to the {\it global} variational problem of minimizing the integral $
\int d^4x A_{\mu}A^{\mu} $.  The gauge fixing condition in its global
form can easily be transformed into  a lattice relation, in terms of
\umu :
\begin{equation}
F(G) := Re\ Tr\ \sum_{\mu} \sum_x \umu^{(G)}= Extremum.
\end{equation}
Obviously, $ F(G) $ has an upper and a lower bound. Accordingly, there
exist both an absolute minimum and maximum. Notice that both maximizing
and minimizing $F(G)$ will satisfy condition (3). Hence we have at
least two Landau configurations which differ by gauge transformation.
This is due to the fact that, for $SU(3)$ gauge theory, the Landau
condition is not sufficient to fix the gauge completely, which leads
to the notorious phenomenon of Gribov copies\cite{Gribov}.

We consider next a given configuration to be cast into Landau gauge.
This will be achieved by driving $F$ to an extremum, in an iterative
process. For constructing \G, we will follow the algorithms
introduced in refs.\cite{dav}\cite{gusor}, which will be referred to as
Cornell\cite{dav} and Los Alamos\cite{gusor} methods. 

{\bf a. The Cornell Method}

In a convergent scheme, the local gauge transformation \G is expected to
approach unity with increasing iteration number $i$.
The Cornell method therefore starts off from the ansatz
\begin{equation}
G(x) = e^{-i\alpha\sum_{\mu}\partial_{\mu} A^{\mu}(x) },
\end{equation}
which is expanded to first order:
\begin{equation}
G(X) \approx 1-i\alpha\sum_{\mu}\partial_{\mu} A^{\mu}(x).
\end{equation}

This approximation requires a subsequent projection of \G back into the
group space. \amu is given by eq. (1) and the lattice form
of $\partial_{\mu} A^{\mu}(x)$ is
\begin{equation}
\partial_{\mu} A^{\mu}(x) = A_{\mu}(x) - A_{\mu}(x-\hat{\mu}).
\end{equation}
The quantity $ \alpha $ is a  real parameter to be suitably chosen,
depending on  the underlying lattice.

{\bf b. The Los Alamos Method}

One introduces 
\begin{equation}
w(x) = \sum_{\mu} \umu + U^+_{\mu}(x-\hat{\mu})
\end{equation}
and rewrites eq.(3)  as
\begin{eqnarray}
F(G)&=& \frac{1}{2} Re\ Tr \sum_x \sum_{\mu}
U^{(G)}_{\mu}(x)+{U^{(G)}_{\mu}}^+(x-\hat{\mu})\\
    &=:& \frac{1}{2} Re\ Tr \sum_{x \in red\ or\ black} w^{G}(x).
\end{eqnarray}
The basic idea is now to construe \G such that F changes its size
monotonically from one iteration step to the next. This 
construction starts from a
checkerboard (red-black) subdivision of the lattice, with \G being
equal to unity on the red (black) sites at even (odd) iteration steps.
The local gauge transformation now simplifies to
\begin{eqnarray}
U_{\mu}(x)\ & \longrightarrow & G(x).U_{\mu}(x).1 \\
U_{\mu}(x-\hat{\mu})\ & \longrightarrow &
1.U_{\mu}(x-\hat{\mu}).G^+(x)
\end{eqnarray}
and can be carried out in parallel on the entire lattice.
This iteration step can be recast into the simple form
\begin{equation}
w(x)\ \longrightarrow w^G(x) = \ G(x)\ w(x).
\end{equation}
The non-unity part of \G is chosen to be a  projection of $w(x)$
onto the group manifold that obeys
\begin{equation}
Re\ tr\ G(x)w(x) \ge Re\ tr\ w(x).
\end{equation}

Note that the change in $F$ is due to independent local changes in $ w(x). $ 
For this reason one has to interchange the role of red and black points after
each step.

In the following we will study the convergence behaviour of these 
basic methods in conjunction with suitable acceleration techniques, 
implemented on the massively parallel machines CM2 and CM5.
We present results obtained on 8 gauge 
configurations  of a $ 8^4 $ hypercubic lattice, equilibrized 
at $ \beta = 5.7, $ as well as 2 configurations of a $16^4 $ lattice ,
at $\beta = 6.0. $ These lattices were generated using a combination of
heat bath and overrelaxation sweeps. The configurations are separated 
from each other by 1000 sweeps.

\section{Techniques}

\subsection {Reunitarization}
We have seen that proper unitarization is an important feature of
gauge fixing iterative schemes. The Gram-Schmidt orthonormalization 
scheme (GS) works very well for the Cornell method but not at all for 
the Los  Alamos algorithm, as it does not ensure the validity eq. (13). 
This can be achieved by the projection method of maximal trace
(MaxTr), which is based on the Cabbibo-Marinari trick\cite{CaMa}:

The projection $G(x)$ of a 3x3 matrix $w(x)$ onto the $SU(3)$ group is
computed  iteratively with the recursive step 
$$ G_i(x)\longrightarrow G_{i+1}(x)=A_1^i A_2^i A_3^i.$$  
and the initial condition $G_0(x)=w(x)$, where
\[ A^i_1 = \left( \begin{array}{ccc}  
        {{\tilde{G}}^{i^*}_{11}} + \tilde{G}^{i}_{22}  & 
        -\tilde{G}^{i}_{12} + {{\tilde{G}}^{i^*}_{21}} & 0 \\
        {\tilde{G}^{i^*}_{12}} - \tilde{G}^{i}_{21}  &  
         \tilde{G}^{i}_{11} + {{\tilde{G}}^{i^*}_{22}} & 0 \\
                0        &         0          & 1
\end{array} \right) \]
\[ A^i_2 = \left( \begin{array}{ccc}
       {{\tilde{G}}^{i^*}_{11}} + {\tilde{G}}^{i}_{33} &  0  &
       -{\tilde{G}}^{i}_{13} + {{\tilde{G}}^{i^*}_{31}}  \\
                0        &  1  &         0           \\
       {{\tilde{G}}^{i^*}_{13}} - {\tilde{G}}^{i}_{31} &  0  &  
        {\tilde{G}}^{i}_{11} + {{\tilde{G}}^{i^*}_{33}} 
\end{array} \right)   \]
\[ A^i_3 = \left( \begin{array}{ccc}
       1  &          0        &         0          \\
       0  & {{\tilde{G}}^{i^*}_{22}} + {\tilde{G}}^{i}_{33} & 
            -{\tilde{G}}^{i}_{23} + {{\tilde{G}}^{i^*}_{32}} \\
       0  & {{\tilde{G}}^{i^*}_{23}} - {\tilde{G}}^{i}_{32} &  
             {\tilde{G}}^{i}_{22} + {{\tilde{G}}^{i^*}_{33}} 
\end{array} \right).  \]
Denoting, 
\begin{eqnarray*}
   N^i_1=& \sqrt{\mid G^{i^*}_{11}+G^i_{22}\mid^2 +
                \mid G^{i^*}_{21}+G^i_{12}\mid^2}    \\
   N^i_2=& \sqrt{\mid G^{i^*}_{11}+G^i_{33}\mid^2 +
                \mid G^{i^*}_{31}+G^i_{13}\mid^2}    \\
   N^i_3=& \sqrt{\mid G^{i^*}_{22}+G^i_{33}\mid^2 +
                \mid G^{i^*}_{32}+G^i_{23}\mid^2},  
\end{eqnarray*}
the elements ${\tilde{G}}^i_{mn}$ in each matrix $A^i_k$ equal
$G^i_{mn} $, divided by the corresponding scaling factor $ N^i_k$.
 
By construction, $ G_i \in SU(3), i \ge 1 $ and  
$ Tr(G_{i+1}G_0) \ge Tr(G_0) $. We achieve maximal trace
(within our 64 bit machine accuracy) after about $ N \sim 5 - 7$
steps, and use $G_N$ for the projection of $w$.

\subsection {Convergence criteria}

We have to define an appropriate quantity that can serve as a
monitor for the quality of gauge fixing achieved during the iteration
process. Our aim is to minimize the quantity 
$ \delta=\mid extremum(F)-F_i\mid $.
As $ extremum(F) $ is not known during the iterative procedure, we have
to use other quantities than $\delta$ to judge the convergence.
In the literature two possiblities have been considered\footnote{In
\cite{hul}, $extremum(F)$ is appearently estimated by the last $F_N$
where $N$ is the last iterative step. This is however a very
misleading quantity}:

\begin{enumerate}
\item One can use \cite{dav}\cite{mandgp}
\begin{equation} 
\sigma _1=\frac{1}{n_c L^4}\sum Tr\ 
(\partial_{\mu}A^{\mu})(\partial_{\mu}A^{\mu})^+
\end{equation}
as a direct measure of fulfillment of the Landau gauge condition.
$L$ is the lattice size, and, for the $SU(3)$ gauge theory, $n_c =3. $ 
\item As $G_i \rightarrow $  unit operator with  $i \rightarrow \infty$
the average trace of the gauge matrices
\begin{equation}
\sigma _2=1-\frac{1}{n_c L^4}Re\ Tr\ \sum G(x)
\end{equation}
can serve as an alternative monitor for convergence \cite{gusig}
\end{enumerate}
As F is quadratic in all variables, we propose here as a third
possibility to employ the {\it rate of change} in the
iteration step $ F_i \longrightarrow F_{i+1} $
as a criterium for convergence achieved at step $i$ 
\begin{equation} \sigma _3=F_{i+1}-F_i. \end{equation}
Notice  that $F$ behaves monotonically.
\begin{figure}[hbt]
\centerline{
\hfill
\epsfysize=150pt \epsfbox{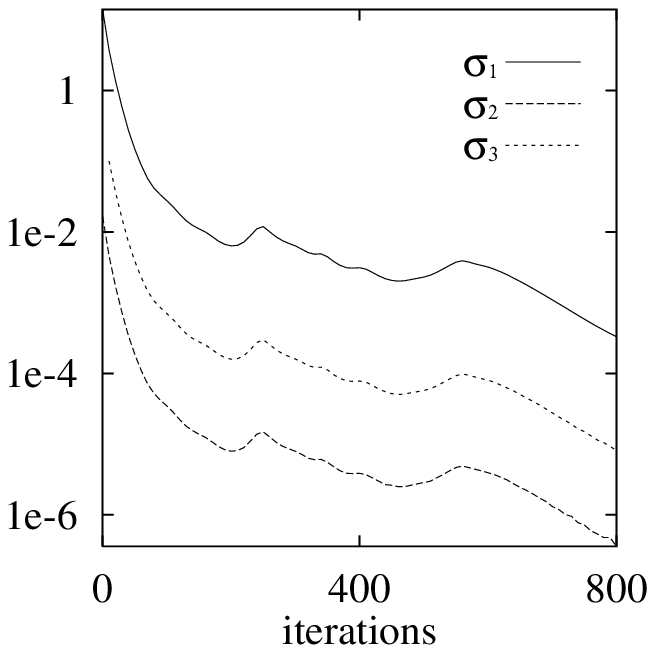}
\hfill
\epsfysize=150pt \epsfbox{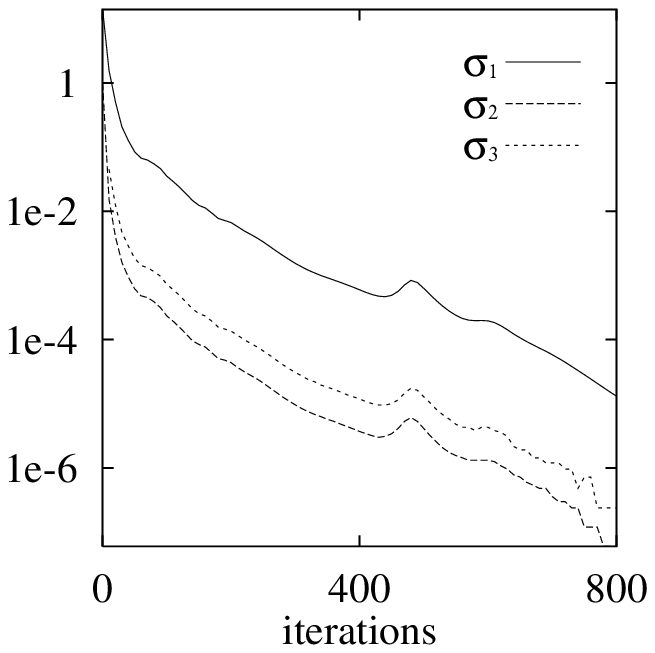}
\hfill
}
\caption{\em \hfill\break
left: The convergence behaviour in maximizing $F,$ using the Cornell
method and reunitarization procedure of maximal trace.  right: The 
same for Los Alamos method.}
\end{figure}
\begin{figure}[hbt]
\centerline{
\hfill
\epsfysize=150pt \epsfbox{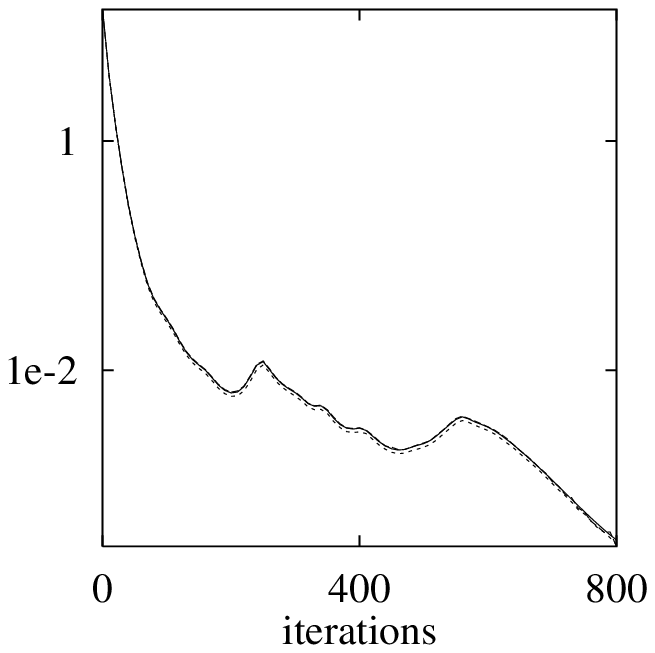}
\hfill
\epsfysize=150pt \epsfbox{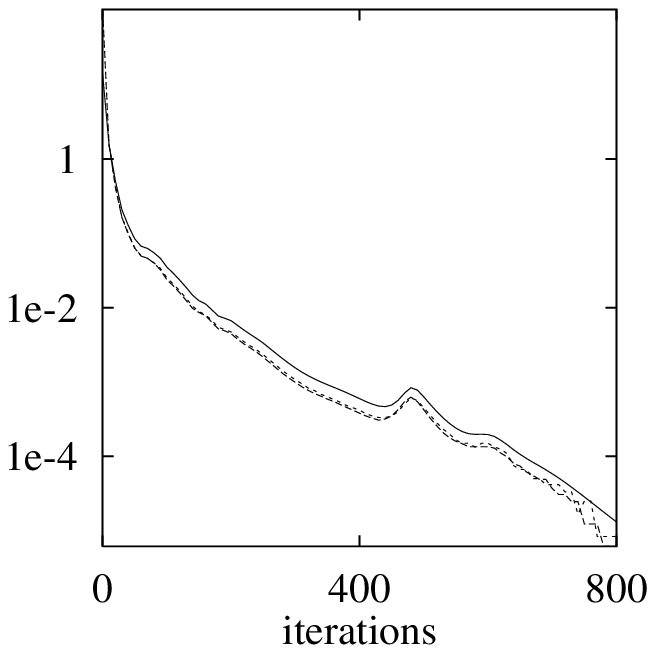}
\hfill
}
\caption{\em Same as in Fig. 1, but with signals appropriately rescaled.
\hspace{6cm}}
\end{figure}

We see from Fig. 1 that the three signals differ considerably in 
size. One must thus be careful when comparing the quality of
gauge fixing quoted by different authors. Nevertheless, 
to the degree that the signals $\sigma _j$ provide useful measures for the
distance $\delta, $ we would expect them to show similar shapes 
during the iterative process, {\it i.e.} coincident positions of
maxima and minima. This is indeed the case, as the signals can nearly
be made to coincide by appropriate rescaling (see Fig.~2).\\

So far, we have considered gauge transformations
for {\it maximizing} $F(G).$  In order to drive the system into a 
{\it minimum}
of $F$,  we simply revert the previous construction, using 
$G^{-1}=G^+$ instead of $G$.
We see in Fig. 3 that the convergence behaviour
of the maximizing and minimizing procedures is very similar
during the first several hundred iterations. After that, the 
minimizing iterative scheme starts to oscillate badly, with poor convergence
compared to the maximizing case. Recall, however, that the minimization
procedure yet rendus a monotonic behaviour in $F$.
\begin{figure}[hbt]
\epsfbox{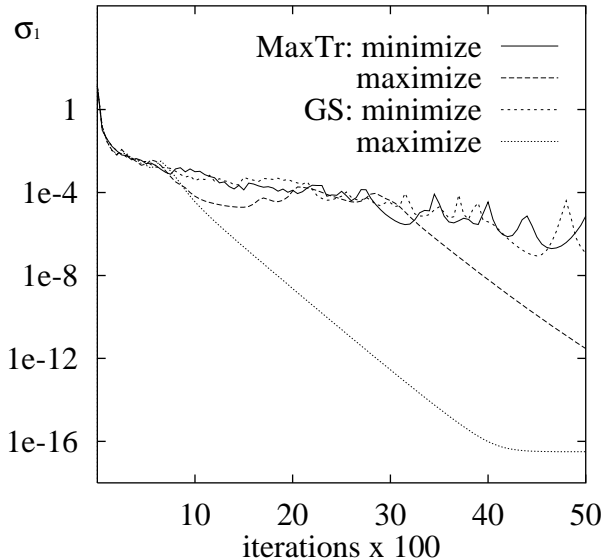}
\caption{\em $\sigma_1$ obtained by maximizing and minimzing $F$ using
different reunitarisation methods. After several hundred iterations, 
it beginns to oscillate widely.}
\end{figure}

\section {Accelerating Procedures}


The discussed algorithms, which are
basically local,  perform sufficiently on small lattices, such as 
$8^4$. On larger lattices, however, they show poor efficiency,
due to the phenomenon of critical slowing down
which occurs when the convergence modulating matrix 
carries a large range of
eigenvalues. Various methods have been proposed
to speed up relaxation, like
preconditioning in the
Fourier space\cite{dav}, overrelaxation\cite{mandor} or multigrid
schemes\cite{gusor}\cite{hul}. 
We want to comment here shortly on some of these methods as
implemented in data-parallel computing.

{\bf Fourier acceleration}\cite{dav} We consider the convergence
behaviour of the Cornell method which it is controlled by the
matrix $ \sum \partial_{\mu} A^{\mu}. $
In momentum space, the modes of this matrix
converge with relaxation times proportional to
$\frac{1}{p^2}$. Thus the overall relaxation $\tau $
time will be determined by the smallest momentum states:
$\tau \simeq 1/p^2_{min} \simeq L^2 $. To overcome this critical
slowing  down, one modifies the gauge transformation matrix \G in
Fourier space such that all modes converge to zero at the same rate
(as fast as the fastest mode):
\begin{equation}
G(x)=e^{-i\alpha W}\ \longrightarrow\ e^{-i\hat{F}^{-1} \left( \frac{a
p^2_{max}}{p^2} \hat{F}\left( W \right) \right)} 
\end{equation}
where $\hat{F}$ denotes the Fourier transform.\\

In Fig.~4 we see that this preconditioning reduces the required number of 
iterations by a factor of about {\it two}\footnote{Nevertheless this
result is not as expected from ref \cite{dav} which quotes a gain
factor of 7 !}.

Note however, that the cost of  the Fourier transformation
is non-negligeable on parallel machines, due
to its wide range communication requirements. We used the fast
Fourier transform (FFT) subroutine, as provided by the
scientific subroutine library (CMSSL) on the CM2. 
This subroutine requires a special ordering of parallel data, 
the so called SEND ordering. Our axis are, however, NEWS-ordered. 
In this case FFT carries out an internal reordering from NEWS to SEND
data structure, prior to the actual computation of FFT\cite{cmssl}.  
As a result, the iteration step on the CM2 is slowed down by
a factor 15, through the two FFT steps, Eq. 17.
\begin{figure}[hbt]
\epsfysize=160pt \epsfbox{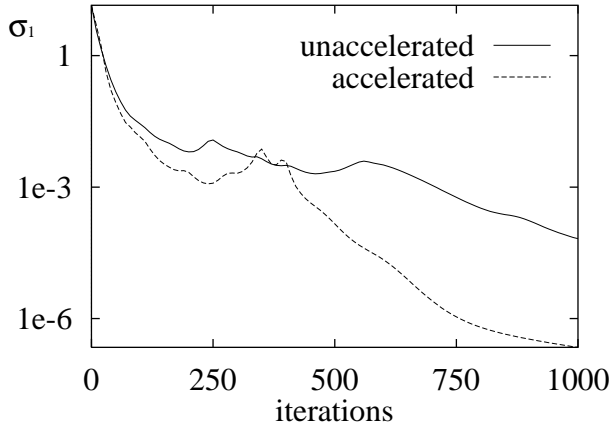}
\caption{\em The effect of Fourier acceleration on the convergence
behaviour for the $8^4$ lattices.}
\end{figure}

{\bf Overrelaxation} 
Overrelaxation methods are widely used in improving convergence of
iterative methods, as well as in overcoming critical slowing down in
Monte Carlo updating. In ref.\cite{mandor} Mandula and Ogilvie applied 
overrelaxation ideas to lattice gauge fixing. They replaced the gauge
transformation matrix \G by an infinite series
\begin{equation}
\tilde{G} = \sum_{n=0}^{\infty} \frac{[\omega]_n}{n!} (G-1)^n,
\end{equation}                                                  
where
\begin{equation}
[\omega]_n = \frac{\Gamma(\omega +1)}{\Gamma(\omega +1-n)}.
\end{equation}
The overrelaxation parameter $\omega $
can take values between 1 and 2. The
optimal choice is to be made empirically\footnote{Approximately, the
optimal choice $\omega_{opt} \sim \frac{2}{1+\frac{3}{L}},$ where L is
the lattice size\cite{hul}\cite{mandor}}.
For our $8^4$ lattices we found that the best value for $\omega$ lies 
near 1.45 (cf. Fig.~5). We truncated the series after two terms (this
corresponds to the original form introduced by Adler \cite{adler}). 
Calculations with higher order terms showed similar behaviour. It is 
worth mentioning that the overhead of the overrelaxation is very small.
\begin{figure}[hbt]
\centerline{
\epsfxsize=250pt \epsfbox{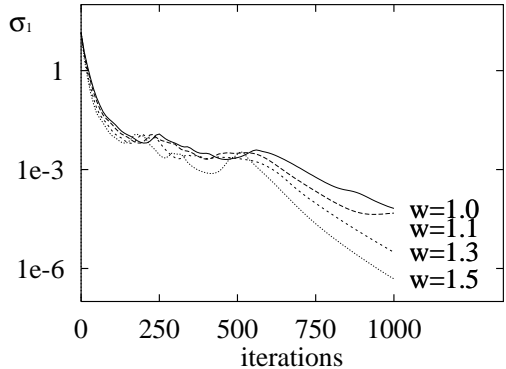}
}
\caption{\em $\sigma_1$ for different overrelaxation parameters.}
\end{figure}

{\bf Stochastic overrelaxation}\cite{gusor}
In this method one applies a local gauge transformation $G(x)^2$ with
probability $p$, instead of always applying \G. For $p=1$ the procedure
definitely diverges. This ``go wrong once in a while'' principle 
has the capability, however, to render a considerable speed up of 
convergence of the two gauging methods treated in this work. 
The actual acceleration gain turns out to depend strongly on $p$.
For our $8^4$ lattices we found  best convergence to occur at
$ p\sim 0.9.$ 
\begin{figure}[bht]
\centerline{
\epsfxsize=250pt \epsfbox{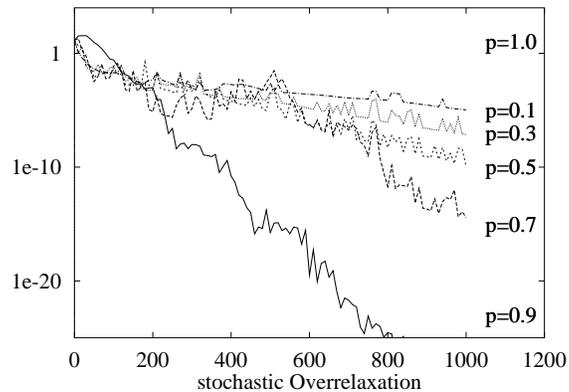}
}
\caption{\em Efficacy of stochastic overrelaxation as a function of the
probability $p$.}
\end{figure}

\section{Performance on CM2 and CM5}

{\bf The Connection Machines in Wuppertal}
The CM2 at the University of Wuppertal is an 8k machine with 256
64-bit floating point accelerators and 256 MByte memory. 
Data IO is performed on a disk parallel
storage system (Data Vault), with 10 Gbyte capacity.
The CM5 is configured as a 32 processor nodes machaine with 4 vector 
units to the  node, and a 16 Gbyte
scalabale disk array (SDA). The CM5 can be used in both message
passing and data parallel
models. Given the 4 dimensional hypercubic structure of our lattices,
we implement the gauging procedures in the data parallel model.

{\bf Implementation}
In the following we will describe some features of our CMFORTRAN 
implementation on the  above CM2 and CM5 machines.\\
First, we have to establish a proper distribution of  our data (gauge 
fields $\{U\}$) among the processors. This has to be chosen to
minimize data traffic.
The data layout has of course a direct influence 
on the subgrid structure, that the CMF compiler produces. 
This is especially important on the CM5 because its performance is heavily
constrained by communication. 
Physicswise, we are dealing with nearest neighbour interactions.
Technically, on the CM5, our nearest neighbour interactions are mapped 
-- by the cshift operation within CMF -- onto
data movements which occur in-processor, on-chip and between-chips,
in decreasing order in speed. It must therefore be the main goal 
to attain a layout that minimizes the amount of off-chip cshift operations. 
That implies a subgrid geometry, for which most cshifts are to be
performed on the longest axis. This is achieved by assigning 
suitable weights to each axis. 

The Cornell method, for example, is obviously isotropic in the sense, 
that all space time directions are equivalent
with respect to the amount of shift operations required.
It is therefore natural to select the ``canonical'' layout for the
gauge field \{$U$\}, with all space time axis declared to be parallel
with equal weights. 
The two matrix (color) indices of \{$U$\} and the Lorentz index $\mu$ are
chosen to be serial, which leaves us finally with
the array structure 
\[ \begin{array}{ll}
U^{(cornell)}=&U(n_c,n_c,n_l,L,L,L,L) \\ 
cmf\$ layout &U(:serial,:serial,:serial,:news,:news,:news,:news).
\end{array} \]
$n_c=3$ is the dimension of color space, $n_l=4$ corresponds to the range
of the Lorentz index and $L$ is the linear size of our hypercubic lattice.

The Los Alamos method, on the other hand, induces an asymmetry into 
the code, due to the red black splitting. We map the entire lattice onto
the red (black) part using a restricted lattice with a geometry 
of the form $(L/2,L,L,L)$. In order to store all the links on such a
lattice we double the range of one particular serial index, which we
choose to be the Lorentz index, $\mu $. Due to the resulting
asymmetric geometry, one expects to enhance the performance 
in assigning appropriate weights to the four parallel axis.
We found that the best performance is achieved by assigning
a relative weight of two to the short axis. As a result, we
work with the array
\[ \begin{array}{ll}
U^{(Los Alamos)}=&U(n_c,n_c,2 n_l,\frac{L}{2},L,L,L) \\
cmf\$ layout &U(:serial,:serial,:serial,2:news,:news,:news,:news).
\end{array} \]

{\bf Performance Data}
For clarity, we restrict ourselves in the
following to quoting measurements from codes produced by the
CMF compiler
(release cmf 2.1 beta 0.1 on the CM5), using complex
arithmetic in double precision. On the CM2, we are running the
slicewise CMF compiler (cmf 1.2).
It goes without saying, that there is room for improvement
of the pure FORTRAN code by resorting to lower level language programming
in some kernel routines\footnote{For $SU(3)$ matrix multiplication, 
e.g., a performance gain of up to 30\% may be reached by programming in 
DPEAC (CDPEAC) on the CM2 (CM5), respectively.}.

We find that the {\it local} $SU(3)$ multiplication has a performance
of about 1.2 Gflops (370 Mflops) on our CM5 (CM2) for a $16^4$ lattice. 
On the CM5, this corresponds to about 30\% of the peak rate 
of 4 Gflops. The reason for this rather low performance lies in the 
fact that  the present CMF compiler does not yet produce optimal
complex arithmetic for the vector units:
it translates a series of complex number multiplications into a code
with too many add-load and mul-load commands, rather than mul-adds.

An important issue is the additional degradation of these
performance characteristics through communication.
Two features have a large impact on this latter loss:
1. The size and shape of the
subgrids residing on the individual processors -- the relevant surface
effects in communication can be influenced by the programmer's layout;
2. the latency of the cshift operations during run time\footnote{The
present implementation of CMF on the CM5 suffers particularly from the large 
latency time of 300 msec.}.
As a result we find, on the CM5, the performance  $P(8^4)$  for the  $8^4$
lattice to be only  65\% of  $P(16^4)$.
For this reason the data presented here refer to $16^4$ lattices only. 

In table 1 we compare performance figures from CM2 and CM5,
as measured on the
Cornell algorithm. Notice that
interprocessor-communication is needed when calculating
the quantity $ \sum \partial_{\mu} A_{\mu} $ according to Eq. 6
configuration $\{U_{\mu}\}$, and when performing the gauge transformation 
$$ U_{\mu}(x)\quad \longrightarrow \quad G(x) U_{\mu}(x)
G(x+\hat{\mu}). $$

In table 2 we present the corresponding performance data from the Los 
Alamos algorithm. Within this method, the gauging step proper is carried out
locally. After each such step, however, one must rearrange 
the links
from red (or black) into black (or red) ordering. This 
gather/scatter stage involves the communication and  deteriorates
the
performance from the pure gauging step, which runs at 1 Gflops
on the CM5.
The communication overhead being nearly 60\% of run-time
on both machines, the floprate is finally degraded to an average 
of 304 Mflops on the CM5, which is merely 8\% of the theoretical
peak performance. Note, that the computing of $\sigma_1$ is more
expensive in this case, due to the data structure.

The situation appears to be more favorable for the Cornell method,
where the time required for communication is 37\% on the CM5, and 27 \%
on the CM2, which leads to an overall performance of 530 MFlops
for the CM5 (198 MFlops for the CM2).\\[1cm]
\begin{tabular}{|l||cc||cc|}
\multicolumn{1}{c}{} & \multicolumn{2}{c}{CM2}
&\multicolumn{2}{c}{CM5} \\
\hline
Subroutine  & MFlops & time in \% & MFlops & time in \% \\
\hline\hline
$\sum \partial_{\mu} A_{\mu}$  & 166.5 & 18.45 & 409.7  & 20.1 \\
$\sigma_1$, $\sigma_2$, $\sigma_3$   & 556.8 &  1.1  & 1481.8 & 1.1  \\
{\it SU(3)-projection}         & 301.6 &  45   &  918.5 & 39.5 \\
{\it transformation}           &  68   & 34.75 & 163.1  & 38.8  \\
\hline
\end{tabular}\\[1.5ex]
{\em Table 1: Benchmarking the Cornell Method }\\[0.2cm]

\begin{tabular}{|l||cc||cc|}
\multicolumn{1}{c}{} & \multicolumn{2}{c}{CM2} &\multicolumn{2}{c}{CM5} \\
\hline
Subroutine  & MFlops & time in \% & MFlops & time in \% \\
\hline\hline
{\it Computation of G}             & 343.85 & 16.6 & 1026  & 16.5 \\
$\sigma_1$                         &  84.5  & 17.6 &  210  &  20  \\
$\sigma_2$, $\sigma_3$                 &  256.3 & 0.4  & 579.8 & 0.55 \\
{\it transformation}           &   367  & 8.86 & 1202  & 7.8  \\
\hline
\end{tabular}\\[1.5ex]
{\em Table 2: Benchmarking the Los Alamos Method }\\[0.1cm]

We should mention, that the results presented in this paper
are based upon a test version of the software
where the emphasis was on providing functionality and the
tools necessary to begin testing the CM5 with vector units.
The CM5 software release has not had the benefit of optimization
or performance tuning and, consequently, is not necessarily
representative of the performance of the full version of this
software.
\vfill
\newpage

{\bf Acknowledgements} We are grateful to R. Flesch, G. Siegert, and
P. Ueberholz for many discussions, and letting us share their
insight of the two CM-systems. We are also grateful to G.S. Bali for
providing us with the gauge configurations. This work was supported by
grant Schi 257/1-4 of Deutsche Forschungsgemeinschaft.


\end{document}